# Spin noise spectroscopy of a single-quantum-well microcavity

S.V. Poltavtsev,[1] I.I. Ryzhov,[1] M.M. Glazov,[1,2] G.G. Kozlov,[1] V.S. Zapasskii,[1]
A.V. Kavokin,[1,3] P.G. Lagoudakis,[3] D.S. Smirnov,[2] and E.L. Ivchenko[2]

[1]*Spin Optics Laboratory, St.-Petersburg State University, 1 Ul'anovskaya, Peterhof, St.-Petersburg 198504, Russia*
[2]*Ioffe Physical-Technical Institute of the RAS, 26 Polytekhnicheskaya, St.-Petersburg 194021, Russia*
[3]*School of Physics and Astronomy, University of Southampton, SO17 1 BJ, Southampton, UK*

We report on the first experimental observation of spin noise in a single semiconductor quantum well embedded into a microcavity. The great cavity-enhanced sensitivity to fluctuations of optical anisotropy has allowed us to measure the Kerr rotation and ellipticity noise spectra in the strong coupling regime. The spin noise spectra clearly show two resonant features: a conventional magneto-resonant component shifting towards higher frequencies with magnetic field and an unusual "nonmagnetic" component centered at zero frequency and getting suppressed with increasing magnetic field. We attribute the first of them to the Larmor precession of free electron spins, while the second one being presumably due to hyperfine electron-nuclei spin interactions.

*Introduction.* In the present-day physics of semiconductor nanostructures, a considerable interest is shown for the fundamental spin-related properties which are also promising in applications. Among optical methods of spin dynamics studies, an important place is given to the Faraday-rotation-based spin noise spectroscopy (SNS) which became well-known and popular during the last several years [1]. The advantages of SNS are primarily owed to its nonperturbative nature because probing the sample response by a weak laser beam in the region of transparency does not lead to any real electronic transitions. Extreme smallness of the magnetization fluctuations detected with the SNS technique calls for the highest polarimetric sensitivity which is achieved by using various electronic or optical means. A real breakthrough occurred when the fast-Fourier-transform (FFT) spectrum analyzers were applied in electronics of the SNS technique [2]. The most straightforward optical way to enhance the polarimetric sensitivity implies increasing intensity of the probe light beam and, simultaneously, leaving the input power of photodetector on the admissible level. This can be implemented either by using high-extinction polarization geometries [3] or by placing the sample inside a high-Q optical cavity [4]. In both cases, the light power density on the sample can be increased by a few orders of magnitude, with the light power on the photodetector and, therefore, the photocurrent shot noise remaining on the same low level.

For low-dimensional semiconductor structures (quantum wells, wires and dots) the problem of polarimetric sensitivity is especially topical. In Ref. [5], in order to increase the signal, the spin noise spectra of $n$-doped GaAs quantum wells were studied in the samples containing ten identical quantum wells (QWs). The measurement of the spin noise spectrum of a layer of InAs/GaAs quantum dots (QDs) in a high-finesse microcavity allowed Dahbashi et al. [6] to perform unique investigation of spin dynamics of a single heavy hole localized in a selected QD. We are not aware of any experimental study of spin noise in a single quantum well.

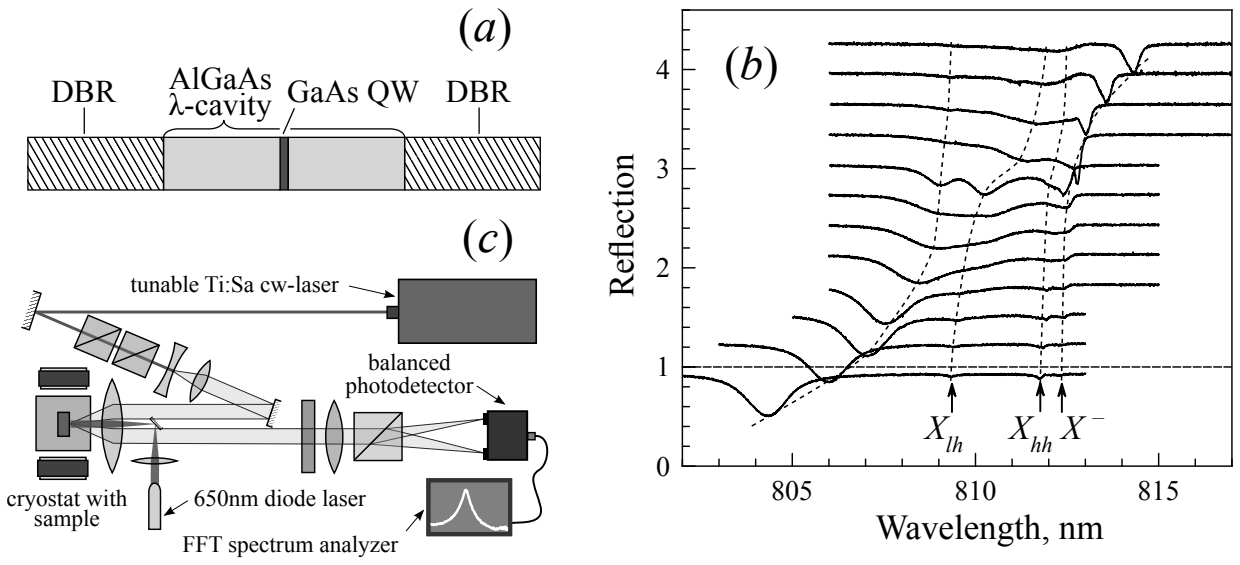

Figure 1: (Color online) (a) Schematic of the sample. (b) Refection spectra measured at different points of the sample, i.e. at different detunings. Arrows mark the positions of negative trion ($X^-$), heavy hole ($X_{hh}$) and light hole ($X_{lh}$) resonances. Different curves are shifted along vertical axis for clarity. Dotted lines are guides for eye and demonstrate anticrossings. (c) Schematic of the experimental setup.

In this paper, we report on the first observation of spin noise in a single GaAs QW embedded inside a high-finesse microcavity operating in the strong coupling regime. A dramatic increase of the sensitivity has made it possible to observe, in addition to the Kerr rotation fluctuations, the noise of ellipticity, the effect reported previously for atomic gases only [7]. We demonstrate also that an increase of the probe beam intensity from weak to moderate values significantly perturbs the spin system in the microcavity making it possible to study the spin noise in steady nonequilibrium states as well [8–10].

*Experiment.* The sample under study represents a 20 nm GaAs QW with AlAs barriers grown along $z \parallel [001]$ axis, placed into the $\lambda$-cavity formed by two distributed Bragg mirrors (DBR) comprised of 25 and 15 pairs of AlAs/AlGaAs layers. Two additional narrow 2.6-nm QWs were grown on both sides of the central well which enabled us to use photodoping by means of the above-barrier illumination. The sample had a gradient of thickness that made it possible to vary the detuning by moving the light-spot on the sample. In more detail, the structure is described in [11]. The schematic of the sample and its reflection spectra are presented in Fig. 1(a) and Fig. 1(b), respectively. The reflection spectra under our experimental conditions of cw excitation were somewhat smoother than those presented in Ref. [11], but still allowed one to trace anticrossings of the cavity mode with material excitations of the QW, namely, the negatively charged trion ($X^-$), heavy- ($X_{hh}$) and light-hole ($X_{lh}$) excitons. The observation of trion resonance means electron density $n_e$ not to be higher than $\sim 5\times10^{10}$ cm$^{-2}$, see Ref. [11] for details. The sample was placed at a temperature of about 6 K into a small transverse magnetic field $B = 0\ldots30$ mT (Voigt geometry), and the fluctuations of the polarization plane rotation were detected in the reflection geometry (Kerr rotation noise) using a standard setup with a balanced photoreceiver (bandwidth 200 MHz) and an FFT spectrum analyzer, see Ref. [1] for details. The signal of ellipticity noise was also measured by placing a properly oriented quarter-wave plate in front of the balanced detector. The probe light from a tunable cw Ti:sapphire laser was tightly focused on the sample (diameter of the spot $\sim$20 $\mu$m) and tuned to the cavity resonance at the chosen point of the sample. In some cases, the probed area of the sample was additionally illuminated by a laser diode with shorter wavelength $\sim$ 650 nm and power density about 25 mW/cm$^2$. Schematic of the experimental set-up is shown in Fig. 1(c).

Under above experimental conditions, in most cases, the Kerr rotation and ellipticity noise were comparable to or even exceeded the shot noise level, so that the noise signals could be easily detected. At the same time, these signals were spatially inhomogeneous with a typical length scale of about 100 $\mu$m. Specifically, depending on particular area of the sample the spin noise signals could be also observed in the absence, rather than only in the persence, of the additional short-wavelength illumination. In this communication, we restrict ourselves to systematic results obtained in our studies of spin noise (SN) spectra of the system at the negative photon-exciton detuning, with the cavity mode lying below the exciton and trion resonances. The dependence of the signals on the magnetic field and probe power was similar in different spots of the sample.

*Experimental results.* Figure 2 demonstrates Kerr rotation noise (a) and ellipticity noise (b) at negative detuning from the $X_{hh}$ resonance $\delta \approx -2.8$ meV. At the chosen point, the noise was observed only under additional illumination. The noise spectrum has been found to contain generally two resonant features with essentially different sensitivity to the applied magnetic field. The frequency of one of them, as expected for a spin resonance, linearly varied with the field (this component is termed "magnetic" hereafter), while the other peak centered at zero frequency did not exhibit any shift with the applied field ("nonmagnetic" component). As seen from Fig. 2, the "nonmagnetic" feature decreases in amplitude with increasing magnetic field. Moreover, the amplitudes and widths of both components depend strongly on the probe beam intensity, as illustrated in Fig. 3. Particularly, with the decrease of the probe intensity, both "magnetic" and "nonmagnetic" resonances narrow down and the relative magnitude of the "magnetic" resonance increases making it possible to observe the field-dependent component of spin noise in the pure form.

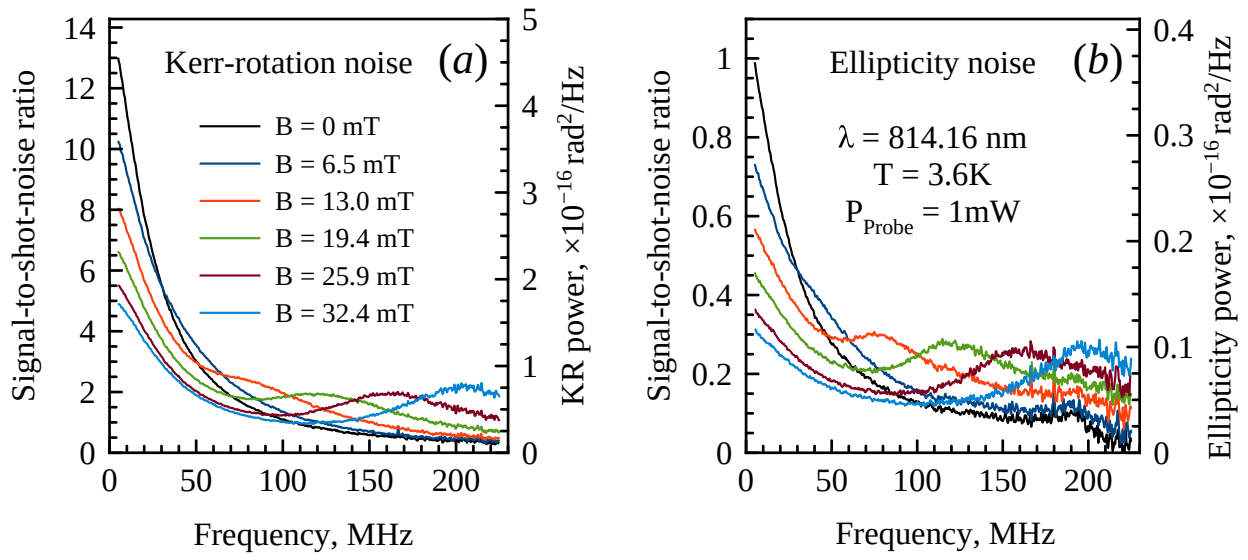

Figure 2: (Color online) Kerr-rotation (a) and ellipticity (b) noise spectra measured at fixed probe power and different magnetic fields indicated at the legend. Shot noise is substracted from the data and the signals are normalized to the shot noise level (see [19] for details of the normalization). Right-hand axis shows the Kerr rotation noise power in rad$^2$/Hz.

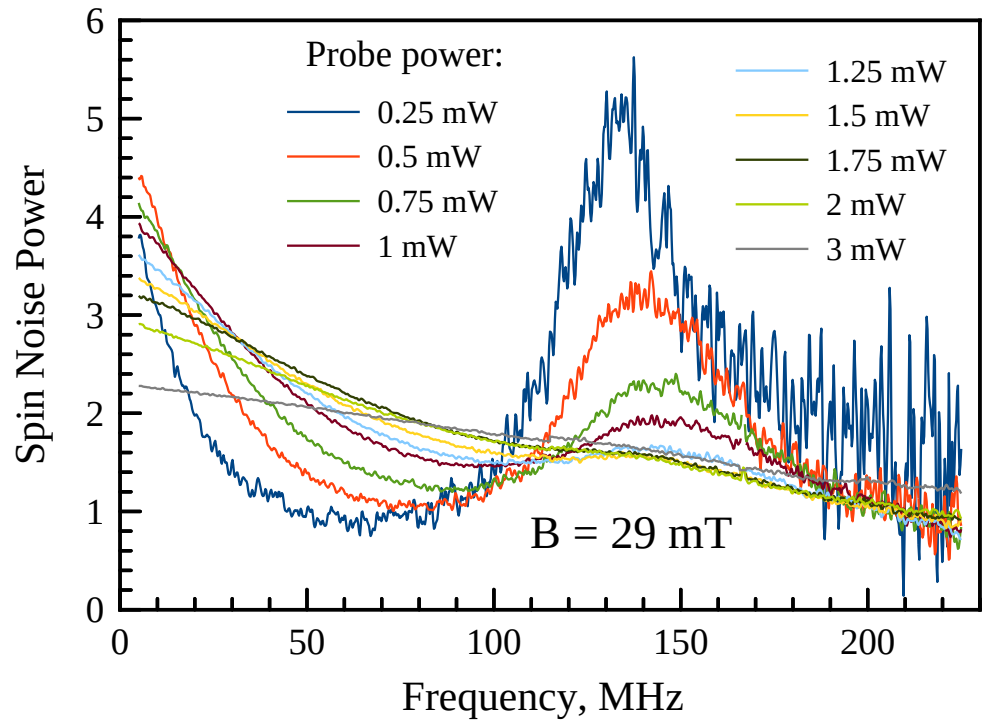

Figure 3: (Color online) Spin noise power density extracted from total Kerr-rotation noise by normalizing signal-to-noise ratio by the probe power measured at $B = 29$ mT and $T = 3.6$ K.

Figure 4(a) presents the Kerr rotation noise spectra at different transverse magnetic fields measured without above-barrier illumination at the sample point where the magnetic component is most pronounced. A field-induced shift of the "magnetic" component corresponded to the effective $g$-factor equal to $|g| \approx 0.33$, which correlates with the electron $g$-factor value in the 20 nm GaAs QW [12]. The shape of this resonance can be well approximated by a field-independent Lorentzian with a FWHM of 60 MHz corresponding to the dephasing time of 6 ns.

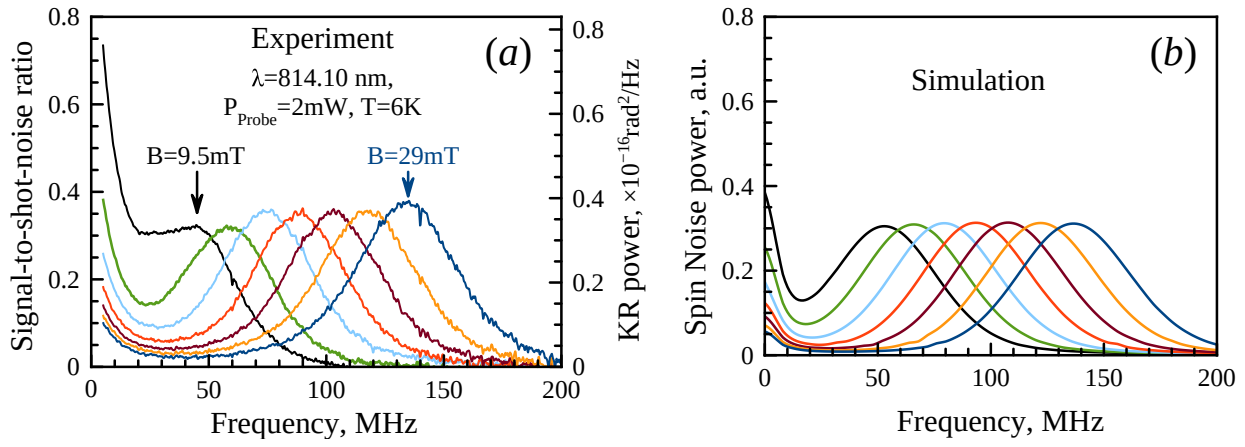

Figure 4: (Color online) (a) Measured Kerr-rotation noise spectra for the magnetic field varied from 9.5 to 29 mT in equal steps. Parameters of the experiment are given in the panel. (b) Calculated spin noise power spectra for $g = -0.33$, $\tau_s \approx 24$ ns, $\delta_e \approx 1.9 \times 10^8$ s$^{-1}$($\approx 30$ MHz). The 7% spread of electron $g$-factor values is taken into account, see [13, 19] for details.

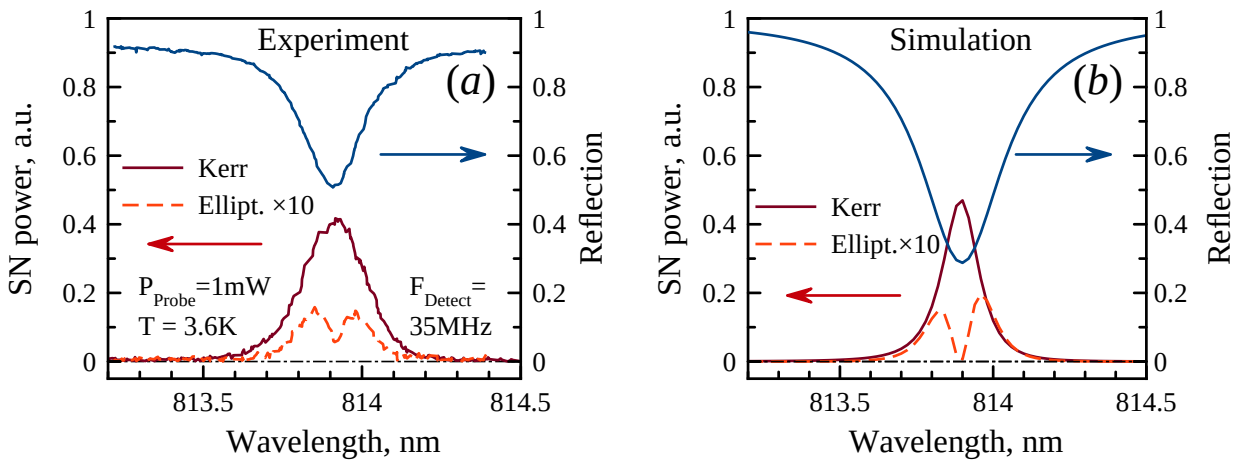

Figure 5: (Color online) (a) Reflectivity (top blue curve, right axis), Kerr rotation (solid/dark red) and ellipticity (dashed/red) optical spectra. (b) Results of calculation after Eq. (1) for $\hbar\omega_c = 1523.52$ meV (wavelength 813.8 nm), $\hbar\varkappa_1 = 0.5$ meV, $\varkappa_2 = 0.14$ meV, taking into account only $X^-$ resonance with $\hbar\omega_{X^-} = 1526.9$ meV (812 nm), $\hbar g_{X^-} = 0.8$ meV, $\hbar\gamma_{X^-} = 0.1$ meV, and taking into account the inhomogeneous broadening 0.8 meV of the trion resonance [19].

The narrow peak at zero frequency can be attributed to the hyperfine interaction with lattice nuclei [13]. Its width of about 15 MHz corresponds to the spin relaxation time $\tau_s = 25$ ns. Overall, such a behavior of the experimental data is well reproduced theoretically, as shown in Fig. 4(b), see below for details.

*Discussion.* The noise of Kerr rotation and ellipticity is caused by the fluctuations of reflection coefficients $r_\pm$ of the microcavity for right (+) and left (−) circularly polarized components of the probe beam. If the probe frequency $\omega$ is close to the cavity resonance frequency $\omega_c$, the reflection coefficients can be presented as [14, 15]

$$r_\pm = -1 + \frac{\mathrm{i}\varkappa_1}{\omega - \omega_c + \mathrm{i}\frac{\varkappa_1+\varkappa_2}{2} + \sum_j \frac{g_{j,\pm}^2}{\omega-\omega_{j,\pm}+\mathrm{i}\gamma_{j,\pm}}}. \quad (1)$$

Here, $\varkappa_1$ and $\varkappa_2$ are the photon escape rates through the mirrors (light is incident on the mirror characterized by $\varkappa_1$), $j$ enumerates resonances in the active layer, namely, $X^-$ trion and $X_{hh}$, $X_{lh}$ excitons, $\omega_{j,\pm}$ are the corresponding resonance frequencies, $g_{j,\pm}$ and $\gamma_{j,\pm}$ are the coupling constants and damping rates, respectively. In general, the differences $\omega_{j,+}-\omega_{j,-}$, $g_{j,+}-g_{j,-}$, $\gamma_{j,+}-\gamma_{j,-}$ are proportional to the $z$-component of magnetization in the system, making the instant values $r_+$ and $r_-$ different [16]. As follows from Eq. (1), the reflection coefficient as a function of the probe frequency has dips at the resonant frequencies of mixed modes, or polaritons [Fig. 1(b)]. We assume that the main contribution to the Kerr rotation and ellipticity fluctuations results from the spin noise of resident electrons and, thus, take into account only trion resonance. In this case, the fluctuations of the trion oscillator strength cause the fluctuating splitting of polariton resonance for $\sigma^+$ and $\sigma^-$ polarizations. As a result the ellipticity noise $\propto |r_+|^2 - |r_-|^2$ should reveal two peaks at the slopes of the resonance and vanish in its center, where $|r_+|^2 = |r_-|^2$, while Kerr rotation noise governed by the phase of the reflection coefficient should be peaked at resonance center. This is demonstrated in Fig. 5 where the experimental data and calculated optical spectra are shown in panels (a) and (b), respectively, see [19] for details.

Strong sensitivity of the spin noise spectra on the probe intensity, particularly, the effects of probe light on amplitudes and widths of the "magnetic" and "nonmagnetic" components in the spin noise spectra, clearly demonstrates that in the strongly-coupled quantum microcavity even a moderate probe perturbs the system. Such a nonequilibrium system calls for special theoretical treatment. The unambiguous presence of the "magnetic" component demonstrates that the noise of Kerr rotation and ellipticity can be attributed to the spin fluctuations of resident electrons, which can be present in the structure due to unintentional doping and/or above-barrier illumination. For relatively low electron densities, $n_e{\sim}10^{10}$ cm$^{-2}$, the carriers are localized at QW imperfections and their spins are affected by both the external magnetic field $\boldsymbol{B}$ and the nuclear field fluctuations. In the strong couling regime, the probe beam, even detuned from material resonances, generates exciton-polaritons and trion-polaritons in the structure. Here we consider the simplest model which takes into account (i) the precession of a localized electron spin in the nuclear field fluctuation with the frequency $\boldsymbol{\Omega}_N$ which is randomly distributed as $\mathcal{F}(\boldsymbol{\Omega}_N) = (\sqrt{\pi}\delta_e)^3 \exp\left(-\Omega_N^2/\delta_e^2\right)$ with $\delta_e$ being the nuclear spin fluctuation [13], (ii) the effect of external magnetic field $\boldsymbol{B}$ with the Larmor frequency $\boldsymbol{\Omega}_B = g\mu_B \boldsymbol{B}/\hbar$, and (iii) probe-induced coupling of electrons and trions neglecting a contribution from excitons. The coupled dynamics of electron and trion spins is described by [16, 17]

$$\frac{\mathrm{d}\boldsymbol{S}}{\mathrm{d}t} = (\boldsymbol{\Omega}_N + \boldsymbol{\Omega}_B) \times \boldsymbol{S} - \frac{\boldsymbol{S}}{\tau_s} - G\boldsymbol{S} + \frac{S_T \boldsymbol{e}_z}{\tau_0^T}, \quad (2a)$$

$$\frac{\mathrm{d}S_T}{\mathrm{d}t} = -\frac{S_T}{\tau_T} + GS_z. \quad (2b)$$

Here $\boldsymbol{S}$ is the electron spin pseudovector with the components $S_x, S_y$ and $S_z$, $S_T$ is the trion pseudospin, $S_T =$

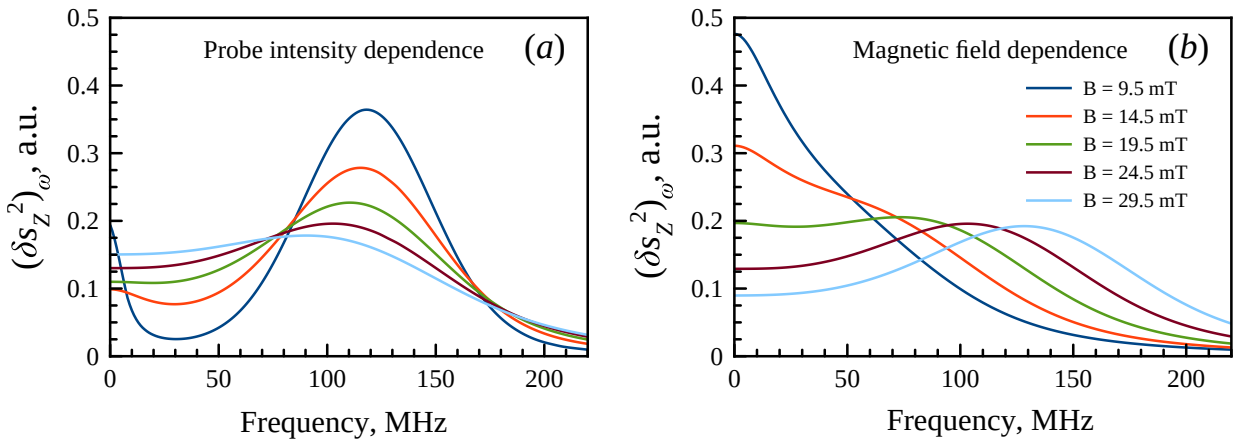

Figure 6: (Color online) Calculated spin noise spectra. (a) Different curves correspond to different trion generation rates $G = 0\ldots 5\times 10^8$ s$^{-1}$ (in equal steps), the magnetic field is fixed, $B = 24$ mT. (b) Different curves correspond to different magnetic fields $B = 9.5\ldots 29.5$ mT (in equal steps), generation rate is fixed, $G = 4\times 10^8$ s$^{-1}$. Other parameters are as follows: $\tau_0^T = 11$ ps, $\tau_T = 9.5$ ps, $\delta_e = 2.5\times 10^8$ s$^{-1}$, spread of $g$-factor values is disregarded.

$(T_+ - T_-)/2$, with $T_\pm$ being the occupation numbers of heavy hole trions with the spin $3/2$ and $-3/2$, respectively, $\boldsymbol{e}_z$ is the unit vector along the growth axis $z$, $\tau_s$ is the electron spin relaxation time and, for simplicity, the effects of spin relaxation anisotropy related to crystallographic orientation of the quantum well are disregarded for simplicity [18], $\tau_0^T$ is the lifetime of the trion, $\tau_T$ is the spin lifetime of the trion given by $\tau_s^T\tau_0^T/(\tau_s^T+\tau_0^T)$ with $\tau_s^T$ being the trion spin relaxation time, and $G$ is the trion generation rate. The latter includes the formation of trions both directly by the probe absorption and via the capture of excitons by resident electrons, it is proportional to the probe intensity and increases with decreasing absolute value of the detuning $|\delta|$. The spin precession in the trion is neglected. We stress that for the linearly polarized probe, the trions thus created contain electrons with any spin orientation, but, according to the optical selection rules, the electron, returned after the trion recombination, has a spin $\boldsymbol{S} = S_T\boldsymbol{e}_z$.

Figure 4(b) demonstrates the calculation of the electron spin noise spectra, $(\delta S_z^2)_\omega$, by using Eqs. (2) in the limit of low probe intensity, $G \to 0$. The parameters of calculation are given in the caption. The model reproduces the main features of measured spin noise spectra, Fig. 4(a): the narrow peak at $\omega = 0$, which vanishes with the increase of the field, and the "magnetic" peak.

An increase in probe intensity and, hence, the trion generation rate $G$ drastically changes the spin noise spectra, as shown in Fig. 6. In qualitative agreement with experimental data presented in Fig. 3, the magnetic peak in the spin noise spectrum decreases, while the peak at $\omega = 0$ becomes broader and relatively more pronounced. Such a behavior can be qualitatively understood bearing in mind that the probe-induced coupling of the electron with trion leads to the anisotropic spin relaxation of the electron. Indeed, at $\tau_s^T \gg \tau_0^T$, the electron spin $z$ component is conserved, while its in-plane components $S_x$ and $S_y$ vanish after the trion decay. As a result, the electron effective spin relaxation rate $\gamma_{\rm eff}$ increases with the field giving rise to the broadening of the spin noise spectrum [5]. Calculation shows that for $G\tau_T \ll 1$, $\Omega_B < G\tau_T/(2\tau_0^T)$ and in the absence of random nuclear fields [19]

$$\gamma_{\rm eff} = \frac{1}{\bar{T}} - \frac{G}{2}\sqrt{\frac{\tau_T^2}{\tau_0^{T\,2}} - \frac{4\Omega_B^2}{G^2}}, \qquad (3)$$

where $\bar{T}^{-1} = 1/\tau_s + G[1-\tau_T/(2\tau_0^T)]$. It follows then that for small enough magnetic fields the spin noise spectrum is centered at $\omega = 0$ and its width increases quadratically with the magnetic field. For $\Omega_B > G\tau_T/(2\tau_0^T)$, the "magnetic" component in the spin noise spectrum appears. The simulation after Eqs. (2) presented in Figs. 6(a) and 6(b) reproduces well not only the magnetic field dependence of the spin noise spectrum, shown in Fig. 2 and measured at the moderate probe intensity, but also the dependence of the spin noise spectrum on the probe intensity, Fig. 3.

A detailed fitting of experimental data by the developed model needs allowance for other possible sources of the "nonmagnetic" component of the spin noise spectrum, e.g., spin fluctuations of holes (in the generated trions or captured in the sample as a result of above-barrier illumination), spin noise of excitons [20] and exciton-polaritons [9], spin noise of electrons and holes trapped in narrow quantum wells or at the localization centers in the barriers. Additionally, "nonmagnetic" component of the ellipticity noise can result from fluctuations of the off-diagonal component of the background dielectric susceptibility tensor, $\mathrm{Re}\{\varepsilon_{xy}\}$, caused, e.g., by the phonons. To elucidate the contributions of particular mechanisms, the application of magnetic field in the Faraday geometry which enhances hyperfine-interaction-induced zero-frequency peak could be useful [6, 13, 21, 22]. All these effects are, however, beyond the scope of present work and deserve further studies.

*Conclusion.* The electron spin noise in a single QW microcavity operating in the strong coupling regime is observed via the Kerr-rotation and ellipticity fluctuations. The spin noise spectrum contains both a "magnetic" component, with its maximum located at the frequency of Larmor precession of the electron spin around the external magnetic field, and a "nonmagnetic" one centered at zero frequency. The magnitudes and widths of these components strongly depend on the probe intensity. The experimental findings are described in the framework of proposed model which takes into account the spin precession of resident electrons in the external magnetic field and the field of nuclear fluctuations as well as the effect of trion generation by the probe beam.

*Acknowledgments.* The financial support from the Russian Ministry of Education and Science (Contract No. 11.G34.31.0067 with SPbSU and leading scientist A. V. Kavokin), Dynasty Foundation, RFBR, and EU projects POLAPHEN and SPANGL4Q is acknowledged.

---

[1] V. S. Zapasskii, Adv. Opt. Photon. **5**, 131 (2013).
[2] M. Romer, J. Hubner, and M. Oestreich, Review of Scientific Instruments **78**, 103903 (2007).
[3] P. Glasenapp, A. Greilich, I. I. Ryzhov, V. S. Zapasskii, D. R. Yakovlev, G. G. Kozlov, and M. Bayer, Phys. Rev. B **88**, 165314 (2013).
[4] A. V. Kavokin, M. R. Vladimirova, M. A. Kaliteevski, O. Lyngnes, J. D. Berger, H. M. Gibbs, and G. Khitrova, Phys. Rev. B **56**, 1087 (1997).
[5] G. M. Müller, M. Römer, D. Schuh, W. Wegscheider, J. Hübner, and M. Oestreich, Phys. Rev. Lett. **101**, 206601 (2008).
[6] R. Dahbashi, J. Hübner, F. Berski, K. Pierz, and M. Oestreich, ArXiv e-prints (2013), 1306.3183.
[7] T. Mitsui, Phys. Rev. Lett. **84**, 5292 (2000).
[8] E. L. Ivchenko, Sov. Phys. Solid State **7**, 998 (1974) [Fiz. Tverd. Tela **7**, 1489 (1974)].
[9] M. M. Glazov, M. A. Semina, E. Y. Sherman, and A. V. Kavokin, Phys. Rev. B **88**, 041309 (2013).
[10] F. Li, Y. V. Pershin, V. A. Slipko, and N. A. Sinitsyn, Phys. Rev. Lett. **111**, 067201 (2013).
[11] R. Rapaport, E. Cohen, A. Ron, E. Linder, and L. N. Pfeiffer, Phys. Rev. B **63**, 235310 (2001).
[12] I. A. Yugova, A. Greilich, D. R. Yakovlev, A. A. Kiselev, M. Bayer, V. V. Petrov, Y. K. Dolgikh, D. Reuter, and A. D. Wieck, Phys. Rev. B **75**, 245302 (2007).
[13] M. M. Glazov and E. L. Ivchenko, Phys. Rev. B **86**, 115308 (2012).
[14] D. F. Walls and G. J. Milburn, *Quantum optics* (Springer, 2008), 2nd ed.
[15] C. Y. Hu, A. Young, J. L. O’Brien, W. J. Munro, and J. G. Rarity, Phys. Rev. B **78**, 085307 (2008).
[16] M. M. Glazov, Physics of the Solid State **54**, 1 (2012).
[17] G. V. Astakhov, M. M. Glazov, D. R. Yakovlev, E. A. Zhukov, W. Ossau, L. W. Molenkamp, and M. Bayer, Semiconductor Science and Technology **23**, 114001 (2008).
[18] N.S. Averkiev, L.E. Golub, Phys. Rev. B **60**, 15582 (1999).
[19] See supplemental materials for details.
[20] D.S. Smirnov, M.M. Glazov, E.L. Ivchenko, to be published.
[21] The effect of magnetic field in the Faraday geometry on spin noise spectra was addressed theoretically in Ref. [13] and experimentally in Refs. [6, 22]. Such an experimental geometry is not possible in our setup.
[22] Yan Li, N. Sinitsyn, D. L. Smith, D. Reuter, A. D. Wieck, D. R. Yakovlev, M. Bayer, and S. A. Crooker, Phys. Rev. Lett. **108**, 186603 (2012).

# Supplemental Material to "Spin noise spectroscopy of a single-quantum-well microcavity"

S.V. Poltavtsev,[1] I.I. Ryzhov,[1] M.M. Glazov,[1,2] G.G. Kozlov,[1] V.S. Zapasskii,[1]
A.V. Kavokin,[1,3] P.G. Lagoudakis,[3] D.S. Smirnov,[2] and E.L. Ivchenko[2]

[1] *Spin Optics Laboratory, St.-Petersburg State University, 1 Ul'anovskaya, Peterhof, St.-Petersburg 198504, Russia*
[2] *Ioffe Physical-Technical Institute of the RAS, 26 Polytekhnicheskaya, St.-Petersburg 194021, Russia*
[3] *School of Physics and Astronomy, University of Southampton, SO17 1 BJ, Southampton, UK*

## S1. NONEQUILIBRIUM SPIN NOISE THEORY

The nonequilibrium system of resident electrons and photoexcited $X^-$ trions is considered. We assume that the excitation is unpolarized or linearly polarized, and it does not serve as a source of spin fluctuations. During photoexcitation the resident electrons are captured to trions and the trion recombination serves as a source of electrons. In the absence of magnetic field the rate equations for the number of spin-up/spin-down electrons, $N_{\pm 1/2}$, and corresponding heavy-hole trions, $T_\pm$, read [1]

$$\frac{\mathrm{d}N_{1/2}}{\mathrm{d}t} = -\frac{N_{1/2}-N_{-1/2}}{2\tau_s} - GN_{1/2} + \frac{T_+}{\tau_0^T}, \tag{S1a}$$

$$\frac{\mathrm{d}N_{-1/2}}{\mathrm{d}t} = -\frac{N_{-1/2}-N_{1/2}}{2\tau_s} - GN_{-1/2} + \frac{T_-}{\tau_0^T}, \tag{S1b}$$

$$\frac{\mathrm{d}T_+}{\mathrm{d}t} = -\frac{T_+-T_-}{2\tau_s^T} + GN_{1/2} - \frac{T_+}{\tau_0^T}, \tag{S1c}$$

$$\frac{\mathrm{d}T_-}{\mathrm{d}t} = -\frac{T_--T_+}{2\tau_s^T} + GN_{-1/2} - \frac{T_-}{\tau_0^T}. \tag{S1d}$$

We denote the total number of electrons in the system as $N_e = N_{1/2} + N_{-1/2}$, the total number of trions in the system as $T = T_+ + T_-$ and the number of resident electrons in the absence of pumping as $N_0$. Since a resident electron can remain resident or can be captured to trion one has

$$N_e + T = N_0. \tag{S2}$$

The solution of the set (S1) and (S2) in the steady state gives

$$N_e = \frac{N_0}{1+G\tau_0^T}, \quad T = \frac{N_0 G\tau_0^T}{1+G\tau_0^T}, \tag{S3a}$$

$$N_{1/2} - N_{-1/2} = T_+ - T_- = 0. \tag{S3b}$$

In the presence of transverse magnetic field, $\boldsymbol{B} \parallel x$, Eqs. (S3) holds, while spin fluctuations obey the following set of equations [1, 2]:

$$\frac{\mathrm{d}\delta S_z}{\mathrm{d}t} = \Omega_B \delta S_y - \frac{\delta S_z}{\tau_s} - G\delta S_z + \frac{\delta S_T}{\tau_0^T}, \tag{S4a}$$

$$\frac{\mathrm{d}\delta S_y}{\mathrm{d}t} = -\Omega_B \delta S_z - \frac{\delta S_y}{\tau_s} - G\delta S_y, \tag{S4b}$$

$$\frac{\mathrm{d}\delta S_T}{\mathrm{d}t} = -\frac{\delta S_T}{\tau_T} + G\delta S_z. \tag{S4c}$$

Here we denote as $\boldsymbol{\delta S} = (\delta S_x, \delta S_y, \delta S_z)$ the electron spin pseudovector fluctuation, $\delta S_T$ is the trion pseudospin fluctuation, $\delta S_T = (\delta T_+ - \delta T_-)/2$, and $\tau_T = \tau_s^T\tau_0^T/(\tau_s^T + \tau_0^T)$. The hole-in-trion spin precession is neglected. Set of Eqs. (S4) is equivalent to Eqs. (2) of the main text in the absence of nuclear fields. The inclusion of spin precession in the total field being the sum of external field and the field of nuclear spin fluctuation is straightforward [3].

The correlation functions of spin fluctuations are introduced as follows:

$$\mathcal{C}_{\alpha\beta}(t) = \langle \delta S_\alpha(t)\delta S_\beta(0)\rangle, \quad \alpha,\beta = z, y, T, \tag{S5a}$$

$$\mathcal{C}^{(+)}_{\alpha\beta;\omega} = \int_0^{+\infty} \mathrm{e}^{\mathrm{i}\omega t}\mathcal{C}_{\alpha\beta}(t)\mathrm{d}t, \tag{S5b}$$

$$\mathcal{C}_{\alpha\beta;\omega} = \int_{-\infty}^{+\infty} \mathrm{e}^{\mathrm{i}\omega t}\mathcal{C}_{\alpha\beta}(t)\mathrm{d}t. \tag{S5c}$$

The same-time correlators read

$$\mathcal{C}_{zz}(0) = \mathcal{C}_{yy}(0) = \frac{N_e}{4} = \frac{N_0}{4(1+G\tau_0^T)}, \tag{S6a}$$

$$\mathcal{C}_{TT}(0) = \frac{T}{4} = \frac{N_0 G\tau_0^T}{4(1+G\tau_0^T)}, \tag{S6b}$$

$$\mathcal{C}_{zT}(0) = \mathcal{C}_{Tz}(0) = 0. \tag{S6c}$$

Taking into account the fact, that the correlation functions of fluctuations obey the same kinetic equations as fluctuations and making the Fourier transform of Eqs. (S4) we immediately obtain the solutions for $\mathcal{C}^{(+)}_{\alpha\beta;\omega}$ [4]. Similar procedure allows us to calculate correlation functions in the presence of nuclear fields $\boldsymbol{\Omega}_N$, in which case the correlation functions should be averaged with the appropriate distribution of the nuclear spin precession frequencies [3]. The results of simulation in Fig. 4(b) of the main text include also the effect of spread of the electron $g$-factor values [3]. For the calculation presented in Fig. 6 the spread of electron $g$-factor has been disregarded, but the value of the nuclear field fluctuation has been somewhat increased to obtain the similar width of the magnetic peak.

For the sake of example, let us analyze in detail the contribution $(\delta S_z^2)_\omega = \mathcal{C}_{zz}(t)$ caused by electron spin fluctuations. We focus on the typical case, where $\tau_s \gg \tau_0^T, \tau_T$. First, we consider the case of zero field, $\Omega_B = 0$. In the limit of vanishing trion generation rate $G$, we have standard result

$$(\delta S_z^2)_\omega = \frac{N_0}{2}\frac{\tau_s}{1+\omega^2\tau_s^2}. \tag{S7}$$

If $G\tau_0^T \ll 1$ (but $G\tau_s$ can be on the order of 1) one has

$$(\delta S_z^2)_\omega = \frac{N_0}{2}\frac{\tau_s'}{1+\omega^2{\tau_s'}^2}, \tag{S8}$$

where $\tau_s' = \tau_s/[1+G\tau_s(1-\tau_T/\tau_0^T)] < \tau_s$. Hence, the spin noise spectrum somewhat broadens with an increase of the trion generation rate due to the depolarization of the electron after the trion recombination: Factor $1-\tau_T/\tau_0^T = \tau_0^T/(\tau_0^T+\tau_s^T)$ is important if hole spin relaxation in the trion is fast enough.

Now we address the case where a transverse magnetic field is applied, $\Omega_B\tau_s \sim 1$. It is noteworthy that, owing to the coupling with trion, the electron in-plane spin component $S_y$ decays with the rate $1/\tau_s + G$, while $S_z$ component relaxes with smaller rate $1/\tau_s'$. Hence, the situation of the anisotropic spin relaxation is realised here (similar to considered in Ref. [5–8]). Making use of the solution for the electron spin dynamics of Ref. [7] and introducing the average relaxation rate

$$\frac{1}{\bar{T}} = \frac{1}{\tau_s} + G\left(1 - \frac{\tau_T}{2\tau_0^T}\right), \tag{S9}$$

and effective precession frequency

$$\tilde{\Omega} = \sqrt{\Omega_B^2 - \frac{G^2\tau_T^2}{4{\tau_0^T}^2}}, \tag{S10}$$

we obtain for the electron spin noise power spectrum

$$(\delta S_z^2)_\omega = \frac{N_0\bar{T}^4}{2\tau_s^3}\frac{1}{[1+(\omega-\tilde{\Omega})^2\bar{T}^2][1+(\omega+\tilde{\Omega})^2\bar{T}^2]} \times\left\{\frac{\tau_s}{\tau_s'}\left[\omega^2\tau_s^2 + (1+G\tau_s)^2\right] + (1+G\tau_s)\Omega_B^2\tau_s^2\right\} \tag{S11}$$

In small magnetic fields, where $\Omega_B < G/2$ the effective precession frequency is imaginary, giving rise to the exponential decay of spin despite the presence of magnetic field. The spin decoherence rate (determining the spin noise spectrum width) is given by

$$\gamma_{\rm eff} = \frac{1}{\bar{T}} - \mathrm{i}\tilde{\Omega} = \frac{1}{\bar{T}} - \frac{G}{2}\sqrt{\frac{\tau_T^2}{{\tau_0^T}^2} - \frac{4\Omega_B^2}{G^2}}, \tag{S12}$$

in agreement with Eq. (3) of the main text. For strong enough magnetic field, $\Omega_B > G/2$, the effective spin precession frequency $\tilde{\Omega}$ becomes real, giving rise to the magnetic peak in the spin noise power spectrum. For $\Omega_B \gg G$ one has $\tilde{\Omega} \approx \Omega_B$.

## S2. PROBE FREQUENCY DEPENDENCE OF SPIN SIGNALS

In order to describe the optical spectra of Kerr and ellipticity noise signals, it is important to include inhomogeneous broadening of the trion resonance frequency in Eq. (1) of the main text. The reflection coefficient has the form

$$r_\pm = \tag{S13}$$

$$-1 + \frac{\mathrm{i}\varkappa_1}{\omega - \omega_c + \mathrm{i}\dfrac{\varkappa_1+\varkappa_2}{2} + \int d\omega_0 \mathsf{F}_{\delta\omega}(\omega_0)\dfrac{g_\pm^2}{\omega-\omega_0+\mathrm{i}\gamma}}.$$

where $g_\pm$ are the coupling constants for right and left circular polarizations differing due to the fluctuating numbers of spin-up/spin-down electrons, $\gamma$ is the damping rate. We have assumed that $\omega_0$ is normally distributed with the variance defined by parameter $\delta\omega$ and the corresponding probability density function $\mathsf{F}_{\delta\omega}$. In derivation of Eq. (S13), we made use of the fact that the main contribution to the spin-Kerr and ellipticity signals caused by the trion resonance are given by spin-induced modulation of trion oscillator strength [2]. Here, for simplicity, we used the notations $\omega_0$, $g$ and $\gamma$ instead of $\omega_{X^-}$, $g_{X^-}$ and $\gamma_{X^-}$ used in the main text.

The fluctuating difference of $g_\pm$ causes Kerr rotation of the incident light polarization plane by angle $\theta_\mathcal{K}$, which is defined by [9, 10]

$$\sin(2\theta_\mathcal{K}) = \frac{2\,\mathrm{Im}\,r_+r_-^*}{|r_+|^2+|r_-|^2}. \tag{S14}$$

The ellipticity angle, $\theta_\mathcal{E}$, can be defined in a similar way

$$\sin(2\theta_\mathcal{E}) = \frac{|r_+|^2-|r_-|^2}{|r_+|^2+|r_-|^2}. \tag{S15}$$

Taking into account that reflection coefficients are nearly equal to each other, $r_+ \approx r_- \equiv r$, and the spin signals are very small, $\theta_\mathcal{K}, \theta_\mathcal{E} \ll 1$, one can find that

$$\theta_\mathcal{K} = \frac{\mathrm{Im}\,r^*r'}{2|r^2|}dg, \quad \theta_\mathcal{E} = \frac{\mathrm{Re}\,r^*r'}{2|r^2|}dg, \tag{S16}$$

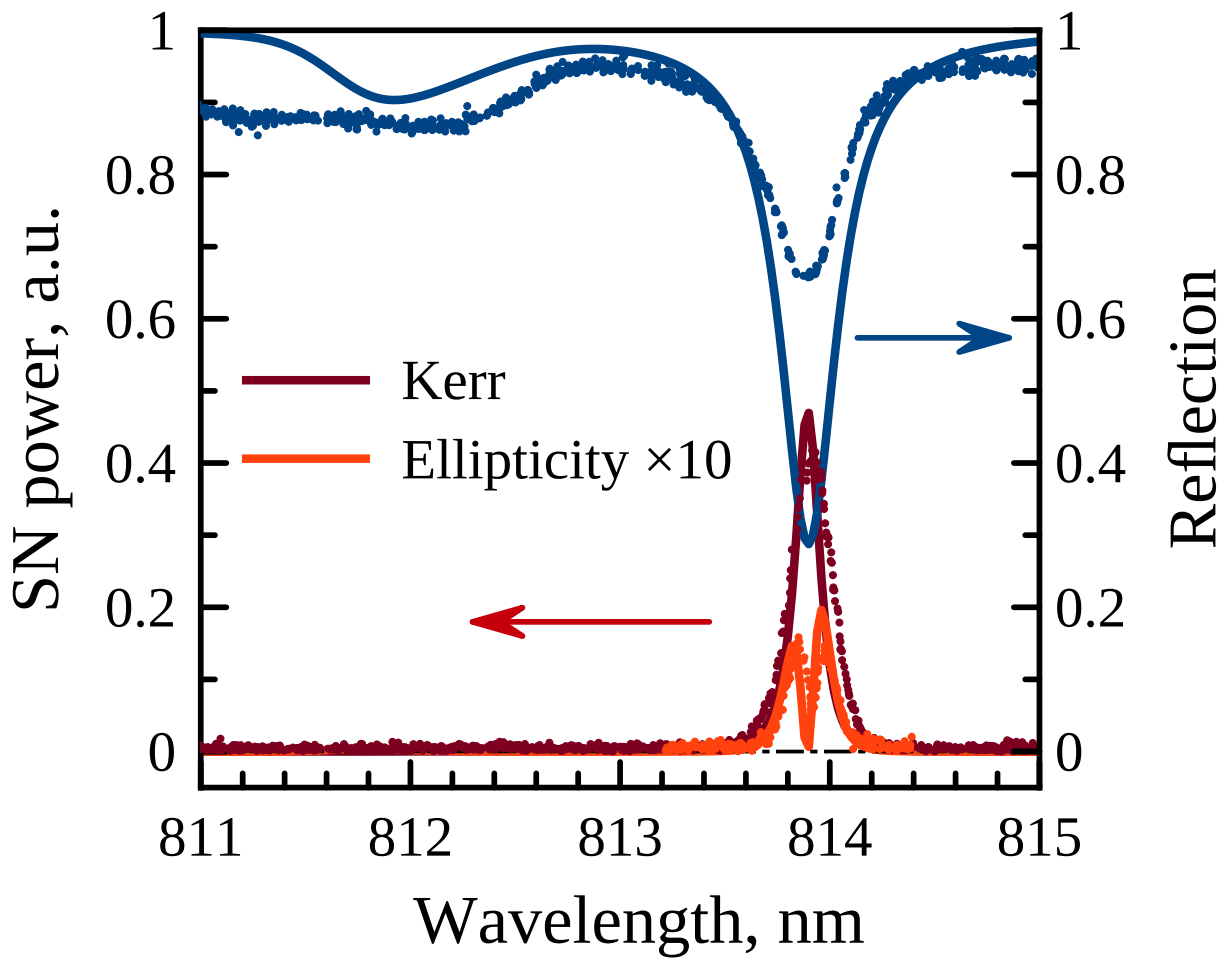

Figure 1: Reflectivity (dark blue), Kerr rotation (brown) and ellipticity (red) optical spectra. Lines present theoretical calculations of $\theta_K^2$ and $\theta_E^2$ after Eqs. (S16), points are the experimental data. The parameters of the calculation are as follows $\hbar\omega_c$ = 1523.52 meV (wavelength 813.8 nm), $\hbar\varkappa_1$ = 0.5 meV, $\hbar\varkappa_2$ = 0.14 meV, $\hbar\omega_0$ = 1526.9 meV (812 nm), $\hbar g = 0.8$ meV, $\hbar\gamma = 0.1$ meV, and inhomogeneous broadening of the trion resonance is 0.8 meV.

where $r' = \mathrm{d}r/\mathrm{d}g$, $dg = g_+ - g_-$.

Qualitative behavior of the Kerr rotation noise and ellipticity noise as functions of the probe frequency could be understood as follows. The momentary fluctuation of resident electron spin results in the splitting of each polariton state into $\sigma^+$ and $\sigma^-$. As a result, the reflection coefficients $r_+$ and $r_-$ have features at different frequencies, $\omega_\pm = \omega_{pol} \pm C\delta S_z$, where $\omega_{pol}$ is the frequency the given polariton state and $C$ is a coefficient. Thus, the dips in $|r_+|^2$ and $|r_-|^2$ are shifted with respect to each other, resulting in vanishing ellipticity exactly at $\omega = \omega_{pol}$. The Kerr rotation is controlled by the phase of the reflection coefficient and has a maximum at $\omega_{pol}$. The frequency dependencies of the reflectivity, Kerr and ellipticity noises calculated after Eqs. (S13),(S16) are presented in Fig. 5(b) of the main text in the vicinity of the cavity mode and in Fig. 1 for the wide range of wavelengths. We stress that for large negative detunings, the inclusion of inhomogeneous broadening of the material (in our case, trion) resonance results in vanishing noise in the vicinity of the trion resonance, while the fluctuations in the vicinity of the cavity mode frequency are observable. The improvement of agreement between the experiment and theory could be reached if one takes into account additional resonances as well as background absorption in the wells and barriers. Such a consideration is beyond the scope of the present work.

Finally, we note that in the presence of photocreated trions, the Kerr (or ellipticity) effect is caused by both electron and trion spins [2], hence,

$$\vartheta = \alpha S_z + \beta S_T, \tag{S17}$$

where $\alpha$ and $\beta$ are coefficients. The correlation function of fluctuations is then given by

$$\langle\delta\vartheta(t)\delta\vartheta(0)\rangle = \alpha^2\langle\delta S_z(t)\delta S_z(0)\rangle + \beta^2\langle\delta S_T(t)\delta S_T(0)\rangle + \alpha\beta\left[\langle\delta S_z(t)\delta S_T(0)\rangle + \langle\delta S_T(t)\delta S_z(0)\rangle\right]. \tag{S18}$$

In addition to the spin noise of electrons and trions this expression contains cross-correlations. However, our estimations show that, for the parameters used to calculate Fig. 4 and Fig. 6 of the main text, the trion and cross-correlation contributions to the observed noise spectra are negligible because the condition $G\tau_0^T \ll 1$ holds.

## S3. NORMALIZATION OF SPIN NOISE SIGNALS

The measured signal at the balanced detector (in Volts) caused by the Kerr rotation or ellipticity is given by

$$\mathcal{B} = A\vartheta P, \tag{S19}$$

where $\vartheta$ is the Kerr rotation angle, $\theta_{\mathcal{K}}$, or ellipticity angle, $\theta_{\mathcal{E}}$ [see Eq. (S16)], $P$ is the probe power, and $A$ is the coefficient dependent on the detector parameters. The autocorrelation function of signal at a detector is given by

$$\langle\mathcal{B}(t)\mathcal{B}(0)\rangle = A^2P^2\langle\vartheta(t)\vartheta(0)\rangle,$$

and the noise power density (in W/Hz) is given by

$$\mathcal{S}_\omega = \frac{1}{R}\int \mathrm{d}t e^{\mathrm{i}\omega t}\langle\mathcal{B}(t)\mathcal{B}(0)\rangle = \frac{A^2P^2}{R}(\vartheta^2)_\omega, \tag{S20}$$

where $R = 50$ Ohm is the input resistance and $(\vartheta^2)_\omega$ is the rotation noise spectrum (in rad$^2$/Hz). The rotation noise power density is quadratic in the probe power $P$.

The shot noise power density scales linearly with the probe power:

$$\mathcal{S}'_\omega = BP, \tag{S21}$$

with the coefficient $B$. Hence, at a given probe power the shot noise is equivalent to the rotation noise with the density

$$(\vartheta'^2)_\omega = \frac{BR}{A^2P}. \tag{S22}$$

In our setup, $A = 2.6\times10^4$ Volt/(rad·W), $B = 1.6\times10^{-12}$ Hz$^{-1}$.

The spin signals presented in Fig. 2(a) and 4(a) of the main text is normalized as follows:

$$\text{Signal-to-shot-noise ratio} = \frac{SN(B) - SN(0.17\ T)}{SN(0.17\ T) - EN}, \tag{S23}$$

where $SN(B) = \mathcal{S}_\omega + \mathcal{S}'_\omega + EN$ is the noise measured at the magnetic field $B$, $EN$ is the noise measured in the absence of probe, i.e. electronic noise.

In Fig. 3 of the main text, in order to elucidate the effect of probe on the spin noise, we additionally divided signal-to-shot-noise ratio by the probe power $P$. In this case, the presented data are proportional to $\mathcal{S}_\omega/P$ or $(\vartheta^2)_\omega$ with a probe power independent coefficient, see Eq. (S20).

[1] G. V. Astakhov, M. M. Glazov, D. R. Yakovlev, E. A. Zhukov, W. Ossau, L. W. Molenkamp and M. Bayer, Semiconductor Science and Technology **23**, 114001 (2008).

[2] E. A. Zhukov, D. R. Yakovlev, M. Bayer, M. M. Glazov, E. L. Ivchenko, G. Karczewski, T. Wojtowicz and J. Kossut, Phys. Rev. B **76**, 205310 (2007).

[3] M. M. Glazov and E. L. Ivchenko, Phys. Rev. B **86**, 115308 (2012).

[4] L. Landau and E. Lifshitz, *Physical Kinetics* (Butterworth-Heinemann, Oxford, 1981).

[5] S. Döhrmann, D. Hagele, J. Rudolph, M. Bichler, D. Schuh and M. Oestreich, Phys. Rev. Lett. **93**, 147405 (2004).

[6] G. M. Müller, M. Römer, D. Schuh, W. Wegscheider, J. Hübner and M. Oestreich, Phys. Rev. Lett. **101**, 206601 (2008).

[7] M. M. Glazov and E. L. Ivchenko, Semiconductors **42**, 951 (2008).

[8] V. Kalevich, B. Zakharchenya, K. Kavokin, A. Petrov, P. Jeune, X. Marie, D. Robart, T. Amand, J. Barrau and M. Brousseau, Physics of the Solid State **39**, 681 (1997).

[9] I. A. Yugova, M. M. Glazov, E. L. Ivchenko, and Al. L. Efros, Phys. Rev. B **80**, 104436 (2009).

[10] M. M. Glazov, Phys. Solid State **54**, 1 (2012) [Fiz. Tverd. Tela **54**, 3 (2012)].